\documentclass[12pt]{article}

\usepackage{epsfig,amssymb}

\def\ltsim{\lower3pt\hbox{$\, \buildrel < \over \sim \, $}}

\def\gtsim{\lower3pt\hbox{$\, \buildrel > \over \sim \, $}}

                                                                                                                                                                                                                                                                                                                                                                                                                                                                                                                                                                                                       \def\be{\begin{equation}}

\def\ee{\end{equation}}

\def\ba{\begin{eqnarray}}

\def\ea{\end{eqnarray}}

\def\ga{\mathrel{\raise.3ex\hbox{$>$\kern-.75em\lower1ex\hbox{$\sim$}}}}

\def\la{\mathrel{\raise.3ex\hbox{$<$\kern-.75em\lower1ex\hbox{$\sim$}}}}

\begin{document}

\baselineskip=16pt

\begin{titlepage}

\begin{center}

\vspace{0.5cm}

\Large{\bf The Effect of 4D Effective Cosmological Constant
On The Stability of Randall-Sundrum Scenario}\\

\vspace{10mm}
Yun-Song Piao$^{a,b}$, Rong-Gen Cai$^{b}$ and Xinmin Zhang$^{a}$ \\

\vspace{6mm}

{\footnotesize{\it
 $^a$Institute of High Energy Physics, Chinese
     Academy of Sciences, P.O. Box 918(4), Beijing 100039, China\\
 $^b$Institute of Theoretical Physics, Chinese Academy of Sciences,
      P.O. Box 2735, Beijing 100080, China \footnote{Mailing address in
      China, Email address: yspiao@itp.ac.cn}\\}}

\vspace*{5mm}

\normalsize

\medskip

\smallskip

\end{center}

\vskip0.6in

\centerline{\large\bf Abstract}

{We study the Rundall-Sundrum model with a small 4D effective cosmological
constant on the brane, and drive a corrected
radion potential following the Goldberger-Wise mechanism. We then discuss the
effect of the 4D effective cosmological constant on the stability of the
brane-system, and find
that to quintessence determined by updated observation, the proper distance
between the two branes required to solve
the hierarchy problem can exist.
However,  during inflation, whether we can get an reasonable
hierarchy scale is still uncertain.
 }

\vspace*{2mm}

\end{titlepage}

As motivated by the string theory \cite{HW} \cite{WL}, there has
been a lot of studies in the recent years on brane models. These
models could provide a solution to problem of the hierarchy
between the Plank scale and the electroweak scale \cite{ADD}
\cite{RS}. In the Randall-Sundrum (RS) \cite{RS} model, the fifth
dimension has orbifold geometry $S_1/Z_2$, and two branes with
opposite brane tension are located at the orbifold fixed points in
a $AdS$ space with negative bulk cosmological constant. The
exponential warp factor in the spacetime metric generates a
hierarchy scale on the observational brane, which is determined by
the distance between two branes. A mechanism of stabilizing this
distance is suggested by Goldberger and Wise (GW) \cite{GW} using
a bulk scalar field.

The existence of a small positive cosmological constant is
strongly indicated by recent observational data \cite{RP}, which
is about $0.7$ of the critical density. This value is $120$ orders
of magnitude less than that from quantum field theory. Some
attempts to get a small cosmological constant in variant of RS
scenarios have been made \cite{CF}. In the meantime, the recent
observations by BooMERANG and MAXIMA \cite{HL} on the location of
the first
peak in the Microwave Background Radiation anisotropy as an strong
evidence of the flat universe strongly support the predictions of
inflation. The corresponding
inflation mechanism in the RS model have been proposed
\cite{PZ}.
Therefore, the consideration of de Sitter phase on the brane is very important,
which is
studied in Ref. \cite{KK,KL}.

In this letter, we study the effect of a small cosmological constant
on the stability of the RS model. We drive the correction of
a small four dimensional (4D)
effective cosmological constant on the 3-brane to the GW radion potential,
and then give some discussions.

We start with a 5D action 
\be S_{bulk} = \int d^4x dy \sqrt{-{g}}
\left({1\over 2\kappa^2} R-\Lambda\right) -\int d^4x
\left(\sqrt{-g_{1}}\Lambda_1+\sqrt{-g_{2}}\Lambda_2\right), 
\ee
where $\kappa$ sets the 5D fundamental scale, $\Lambda$ is the
cosmological constant of the bulk, $g_{1}$ and $g_{2}$ are the
induced metrics on two 3-branes with the corresponding brane
tensions are $\Lambda_1$ and $\Lambda_2$, located at $y=0$ and
$y=r$, respectively. A metric ansatz with maximal symmetry given
by Ref. \cite{KK} is 
\be ds^2=\left(\cosh(ky)-
{\kappa\Lambda_1\over \sqrt{6(-\Lambda)}} \sinh(ky)\right)^2
\left(-dt^2+e^{Ht} d{\bf x}^2\right) +dy^2 , 
\label{ds2} 
\ee The
corresponding matching condition at $y=r$ 
\be
k\left(1-{\kappa^2\Lambda_1\Lambda_2\over 6(-\Lambda)}\right)
\sinh{(kr)}={\kappa^2\over 6}(\Lambda_1+\Lambda_2)\cosh{(kr)}
\label{li} 
\ee 
must be satisfied, where $k=\sqrt{{-\kappa^2
\Lambda\over 6}}$ and 
\be 
H^2 ={\kappa^4 \over 36} \Lambda_1
-{\kappa^2 \over 6}\Lambda \label{h2} 
\ee 
is the Hubble parameter
on two 3-branes. Assuming that the energy densities of two
3-branes are expressed as $\Lambda_1 \rightarrow \sigma
+\Lambda_{1eff}$ and $\Lambda_2 \rightarrow \sigma +
\Lambda_{2eff}$, then fine-tuning $\Lambda = {\kappa^2
\sigma^2\over 6}$, when $\Lambda_{1eff}, \Lambda_{2eff}\ll
\sigma$, which means a very small net cosmological constant on the
3-brane, we can rewrite Eqs. (\ref{li}) and (\ref{h2}) as 
\be
\Lambda_{1eff} = -\Lambda_{2eff} \exp{(-2kr)} \footnote{The
appearance of this limitation is because we assume that the
back-reaction of the bulk scalar field is negligible.} , 
\ee 
\be
H^2 \simeq {\kappa^4\over 18}\sigma \Lambda_{1eff}\equiv
{\kappa^2_4 \Lambda_{eff}\over 3} , 
\ee 
where $\Lambda_{eff}
\simeq {\kappa^4\over 6\kappa_4^2}\sigma \Lambda_{1eff}$ is the 4D
effective cosmological constant on the 3-brane. For convenience,
we define $\eta\equiv \sqrt{{\kappa^2\sigma^2 \over
6\Lambda}}-1\ll 1$, then see that $\eta \simeq {\kappa_4^2\over
\kappa^2 \Lambda} \Lambda_{eff}$ . When $\eta\rightarrow 0$, the
4D effective cosmological constant tend to 0. Therefore, we can
regard $\eta$ as a measurement of 4D effective cosmological
constant.

We can see from the metric (\ref{ds2}) that the distance between two branes
required to solve the hierarchy problem
is less than that in the absence of cosmological constant, but when
$\eta \exp{(2ky)}\simeq 2$,
for $ky\gg 1$, the warp factor have a singularity in finite distance,
thus the other brane ought
to be placed in $\eta \exp{(2ky)}<2$.
Fig. 1 reflects the relation between the location of the other
brane, which is required to generate the hierarchy between the Planck scale and
the electroweak scale, and the 4D effective
cosmological constant on the 3-brane. We see that
with the increasing of the cosmological constant, the location of the other
brane is gradually close to the singularity, but never arrive at it.
Therefore, a model of RS type with the 4D effective cosmological
and the proper distance without the singularity, which is required to generate
the hierarchy, is reasonable.

To study the effect of a small effective cosmological constant
to the stability of this model, following the GW mechanism \cite{GW},
we introduce a bulk scalar field
into the model,
\be
S_{bulkscalar}={1\over 2}\int d^4 x \int d y
\sqrt{-g}\left(g_{MN}\partial^{M}\phi\partial^{N}\phi
-m^2\phi^2\right),
\label{s}
\ee
where $g_{MN}$ is the 5D
metric given in (2), and the boundary potentials of the scalar
field are
\be
S_1=-\lambda_1\int d^4 x\sqrt{-g_1}\left(\phi^2-v_1^2\right)^2~~~%
\mathrm{at}~~~y=0
\ee
\be
S_2=-\lambda_2\int d^4 x\sqrt{-g_2}\left(\phi^2-v_2^2\right)^2~~~%
\mathrm{at}~~~y=r
\ee
where $v_1$ and $v_2$ are the vacuum expectation values of bulk scalar field
on two 3-branes,
$\lambda_1$ and $\lambda_2$ are corresponding coupling constants.

We are interested in those configurations of the bulk scalar field where
the boundary
potentials are minimized. This essentially amounts
to negligible dynamics of $\phi$ along the direction tangential to any of two
3-branes. This assumption is reasonable because we focus on the
stability of two 3-branes system at the moment and do not study
phenomenological consequence of possible coupling of the bulk
scalar field
$\phi$ to matter fields living on two 3-branes. Therefore, it suffices to
concentrate on the equation of motion of $\phi$ only in $y$ direction,
which is
\be
\partial^2_y\phi - 4 f(y)\partial_y\phi - m^2\phi =0 ,
\ee
where
\be
f(y)={k(e^{-ky}+\eta\cosh{(ky)})\over e^{-ky}-\eta\sinh{(ky)}}.
\ee
For $\eta \exp{(2ky)}\ll 1$,
{\it i.e.} a very small cosmological constant, this equation is reduced to
\be
\partial^2_y\phi - 4k(1+\eta e^{2ky})\partial_y\phi - m^2\phi =0 ,
\label{par}
\ee
Limiting
ourselves to
the regime where $\epsilon$ is small,
$\epsilon \equiv \nu - 2 \approx \frac{m^2}{4k^2}\ll 1$, where
$\nu \equiv \sqrt{2+{m^2\over k^2}}$, and only focusing on the
effect of the 4D cosmological constant, we ignore the term of
$\epsilon$ (for the effect of $\epsilon$, see Ref. \cite{YS}), and
in the meantime reserve the first order term of $\eta\exp{(2ky)}$.
Thus the solution of Eq. (\ref{par}) is 
\be 
\phi\simeq a
e^{(2-\nu)ky}+b e^{(2+\nu)ky} +{4\over 3}b (\eta e^{2ky})
e^{(2+\nu)ky}+O((\eta e^{2ky})^2,\epsilon) , 
\label{phi} 
\ee 
where
$a$ and $b$ are two constant determined by appropriate boundary
conditions on two 3-branes. Minimizing the boundary potential at
$y=0$ and $y=r$, we drive 
\be 
a\simeq {v_1 -v_2 (1+{4\over
3}\eta)e^{-(2+\nu) kr}\over 1-(1+{4\over 3}\eta)e^{-2\nu kr}} ,
\label{a} 
\ee 
\be 
b\simeq {-v_1 e^{-2\nu kr}+v_2 e^{-(2+\nu
)kr}\over 1-(1+{4\over 3}\eta )e^{-2\nu kr}} . 
\label{b} 
\ee
Substituting Eqs. (\ref{phi}), (\ref{a}) and (\ref{b}) into the
action (\ref{s}) and making integration of $y$, we get an
effective potential for the radion from the bulk scalar field part
\be 
V_{scalar}(v) \simeq 4 kv^4\left(v_2-v_1
v^{\epsilon}\right)^2+{8\over 3}\eta k v^2 \left(v_2-v_1
v^{\epsilon}\right)^2 , 
\ee 
where $v=\exp{(-kr)}$ denotes the
variability of the distance between two 3-branes. Compared with
the potential in absence of the effective 4D cosmological constant
\cite{GW}, this has an additional correction from a 4D effective
cosmological constant, but we see that this term do not change the
minimum of the effective potential. However, there is still a
contribution from the curvature of the universe in de Sitter brane
\cite{GR}, 
\be 
V_{curvature}={1\over 2k\kappa^2}v^2 R_4, 
\ee 
where
$R_4 =12 H^2$ is the 4D scalar curvature. Consequently, combining
Eq. (16) and (17), the effective potential is reduced to 
\be
V_{eff}(v)=4 k v^4\left(v_2-v_1 v^{\epsilon}\right)^2+{8\over
3}\eta k v^2 \left(v_2-v_1 v^{\epsilon}\right)^2 +{12\over
\kappa^2}k \eta v^2 . 
\ee 
For $\eta\exp{(2kr)} \ll 1$, we plot
Fig. 2, for $k\sim {1\over \kappa} \sim m_p$ and $\exp{(ky)}\simeq
10^{16}$, which is required to generate the hierarchy between the
Plank scale and electroweak scale. We see that a small
cosmological constant can raise the value of the effective
potential in the minimum and make this minimum unstable. When the
cosmological constant increases to certain value, $\eta \sim
10^{-36}$, this minimum disappears, and the correspondent value of
the cosmological constant is $\Lambda_{eff}\sim 10^{-36}m_p^4$ for
$\Lambda\sim {1\over \kappa^5}$.

In summary, we have studied the effect of a very small
cosmological constant on the stability of the RS model and show
that if there is a small cosmological constant on the
observational 3-brane, the distance between two branes required to
solve the hierarchy problem is less than that in the absence of
cosmological constant and without singularity. We give an
effective potential about this distance in some detail, and find
that when the cosmological constant is very small, the
corresponding minimum remain surviving, but is unstable. As the
cosmological constant increases, this minimum disappear, and in
this case we hardly determined the distance between two branes by
this mechanism. For $k\sim {1\over \kappa} \sim m_p$, $\Lambda\sim
m_p^5$ and $\exp{(ky)}\sim 10^{16}$, while the minimum appears,
the cosmological constant value is $\Lambda_{eff}\sim
10^{-36}m_p^4$. We see that to quintessence determined by updated
observation, there exists a minimum, and we can get the proper
distance between two 3-branes required to solve the hierarchy
problem by the GW mechanism. However, to inflation in early
universe, there requires a more high energy scale, in this period
whether we can get an reasonable hierarchy scale is still
uncertain.

\textbf{Acknowledgments}

The authors would like to thank Professor Yuan-Zhong Zhang for
useful discission.
 This work is supported in part by National Natural Science Foundation of
China under Grant Nos. 10047004, and also supported by Ministry of
Science and Technology of China under Grant No. NKBRSF G19990754.

\vspace{5cm}

\begin{figure}[ht]
\begin{center}
\mbox{\epsfig{file=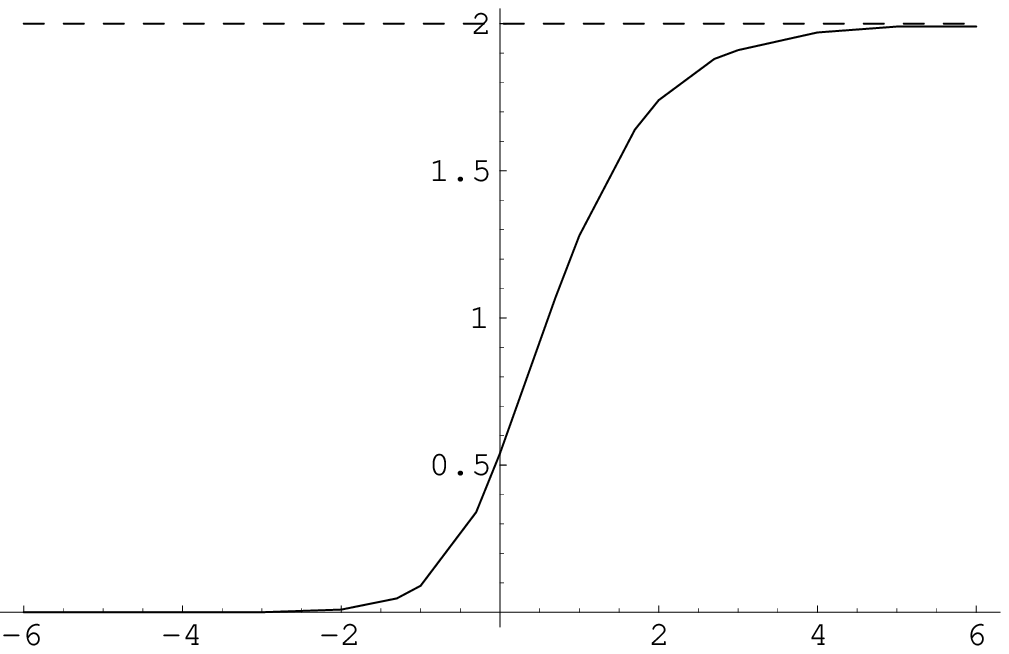,width=8cm}}
\caption {The $y$-axis is ${\eta\over e^{2kr}}$ and the $x$-axis is
$\log_{10}{({\eta\over 10^{-32}})}$. The solid line and the dashing line
denote the location of
the other brane and the singularity, respectively.
}
\end{center}
\end{figure}

\begin{figure}[ht]
\begin{center}
\mbox{\epsfig{file=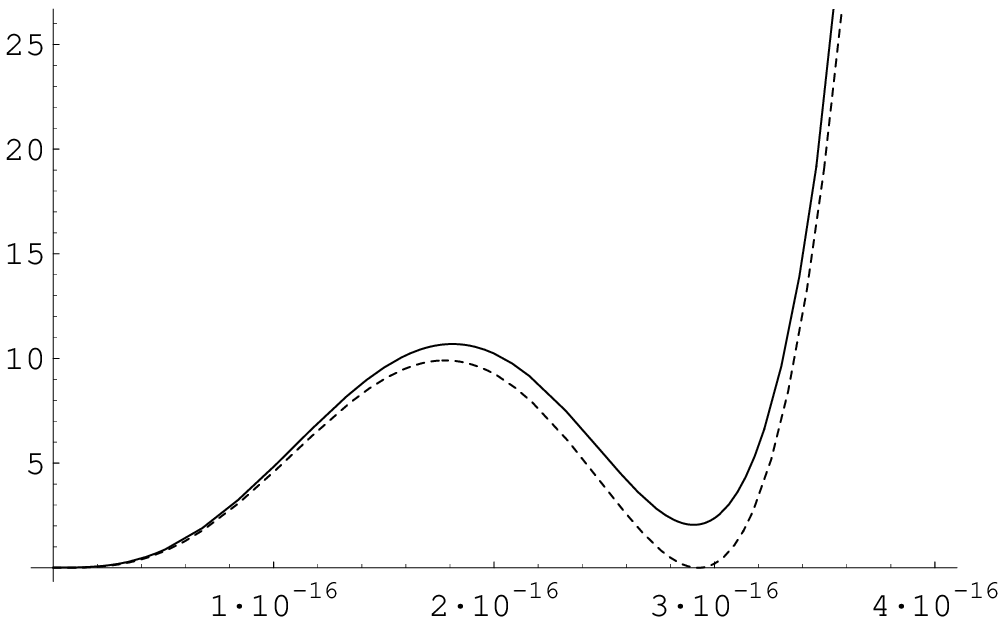,width=6cm}\epsfig{file=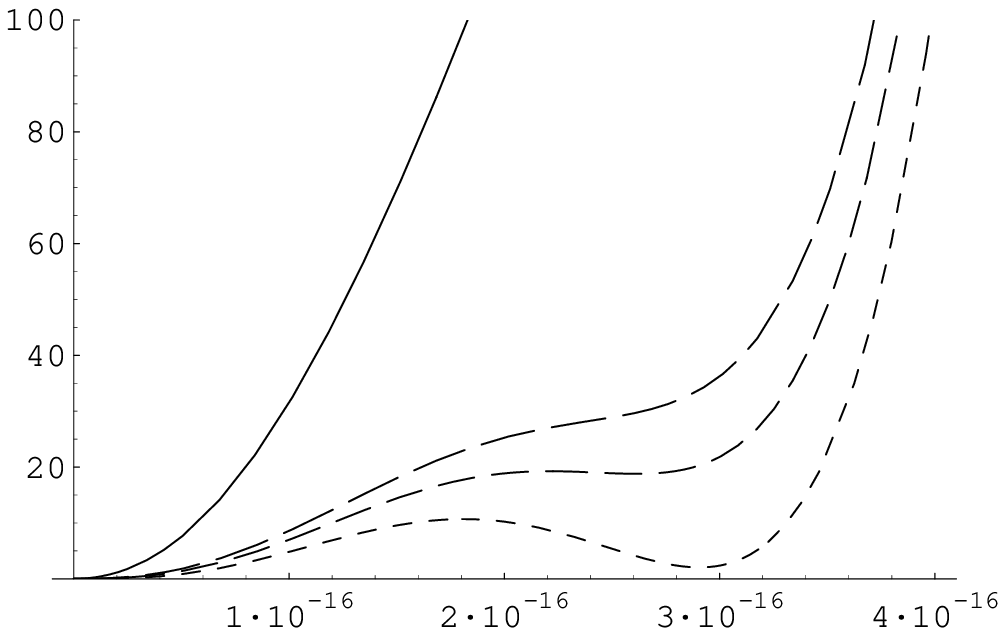,width=6cm}}
\caption {The $y$-axis is $V(\phi)$ and the $x$-axis is $\phi$.
The dashing line of the left figure is that without the 4D cosmological
constant and The right figure reflects the variation of $V(\phi)$
with the increasing of
the 4D cosmological
constant.
}
\end{center}
\end{figure}

\end{document}